\documentclass[a4paper,letter]{hal}
\usepackage{graphicx}
\usepackage{xspace,colortbl}
\usepackage{amsfonts}
\usepackage{amsmath}
\usepackage{amssymb}
\usepackage{supertabular}
\usepackage{rotating}
\usepackage{longtable,geometry}
\usepackage[english]{babel}
\usepackage[utf8]{inputenc}
\usepackage[babel]{csquotes}
\usepackage{listings}
\usepackage[usenames,dvipsnames,svgnames,table]{xcolor}
\usepackage{multicol}
\usepackage{float}
\usepackage{centernot}
\pdfmapfile{=mtpro2.map}
\usepackage[colorlinks=true,linkcolor=Black,urlcolor=MidnightBlue,citecolor=MidnightBlue]{hyperref} 

\newcommand{\JCTCformat}[4]{{\it #1} {\bf #2}, {\it #3}, {#4}.}

\newcommand{\Ref}[4]{\JCTCformat{#1}{#2}{#3}{#4}}

\newcommand{\jcp}[3]{\Ref{J. Chem. Phys.}{#1}{#2}{#3}}
\newcommand{\jpc}[3]{\Ref{J. Phys. Chem.}{#1}{#2}{#3}}

\newcommand{\jmathchem}[3]{\Ref{J. Math. Chem.}{#1}{#2}{#3}}
\newcommand{\jmathphys}[3]{\Ref{J. Math. Phys.}{#1}{#2}{#3}}

\newcommand{\physrev}[3]{\Ref{Phys. Rev.}{#1}{#2}{#3}}

\newcommand{\revmodphys}[3]{\Ref{Rev. Mod. Phys.}{#1}{#2}{#3}}

\begin{document}

\begin{frontmatter}

\title{Generalization of the concepts of seniority number and ionicity}

\author{Thomas Perez, Patrick Cassam-Chena\"{\i}} 

\address{Universit\'e C\^ote d'Azur, LJAD, UMR 7351, 06100 Nice, France}

\begin{abstract}
We present generalized versions of the concepts of seniority number and ionicity. These generalized numbers count respectively  the partially occupied and fully occupied shells for any partition of the orbital space into shells. 
The  Hermitian operators whose eigenspaces correspond to wave functions of definite generalized seniority or ionicity values are introduced. The generalized seniority numbers (GSNs) afford to establish refined hierarchies of configuration interaction (CI) spaces within those of fixed ordinary seniority. 
Such a hierarchy is illustrated on the buckminsterfullerene C$_{60}$ molecule. 
\end{abstract}
\end{frontmatter}


Suggested running head:
General seniority

All correspondance to be send to P. Cassam-Chena\"{\i},\\
cassam@unice.fr,\\
tel.: +33 4 92 07 62 60,\\
fax:  +33 4 93 51 79 74.
%
%
%


%
%
%


\newpage

\section{Introduction}
The concept of seniority originates from the work of Racah on electrons in an atom \cite{Racah1943}. Since then, it has become very popular in nuclear physics \cite{Dean2003,Zelevinsky2003}. Its (re-)introduction in quantum chemistry by Bytautas et al. \cite{Bytautas2011} has proved very fruitful and has inspired many recent works (see \cite{Alcoba2014,Chen2015,Limacher2015,Limacher2016} to quote a few). It is closely related to the notion of ionicity that has been used in valence-bond (VB) theory since the sixties \cite{Simonetta1968}. Sometimes, it is used in some algorithms without actually being named \cite{Karadakov2012,Olsen2015}.\\

Seniority affords to partition the $n$-electron Hilbert space into subspaces spanned by sets of Slater determinants having a definite number of unpaired orbitals. For closed-shell systems, it has been observed that the Full Configuration Interaction (FCI) energy is dominated by the contribution of the seniority-zero part of the wave function, when delocalized molecular orbitals are used, and that, the higher the seniority number of the determinants, the less important their contribution on average \cite{Bytautas2011}. The situation is reversed in VB calculations \cite{Chen2015}.\\

However, even if one restricts a CI space to a subspace of a given seniority number, the size of the CI can be prohibitively large.
Therefore, it is of interest to push the seniority number partitioning strategy a step further, that is to say, to create other such numbers to further break down the seniority-zero subspace into a hierarchy of smaller subspaces. The purpose of this paper is to present a method to define generalized seniority numbers.

Our definition is based on the concepts put forward in Chapter 4 of the Ph. D thesis of M. Vivier, entitled  \emph{``Sur quelques th\'eor\`emes d'alg\`ebre ext\'erieure''}~\cite{Vivier56}, and on their generalization to the case where the shells are not all of the same even dimension. As we shall see, the usual seniority numbers appear in the particular case of primitive shells, hence the term ``generalized seniority'' we have coined for the general case. Our generalization is different from that of Talmi in nuclear structure theory \cite{Talmi71}, where the partitioning is still in terms of pairs of particles, but where the form of the pairing functions goes beyong the simple Slater determinantal one.\\

The paper is organized as follows.
In the first section, the concepts of generalized seniority number (GSN) and generalized ionicity (GI) are defined and explained.
Then, we highlight interesting mathematical results relevant to this concept.
Finally, we illustrate 
how to define a hierarchy based on generalized seniority numbers  on the $\pi$-electron system of the buckminsterfullerene C$_{60}$, and conclude.\\

\section{Generalized seniority number and generalized ionicity}

We consider a one-particle Hilbert space $V$ which is the direct sum (not necessarily orthogonal) of $n$ vector subspaces $V_1,\ldots,V_n$ of respective dimensions $2d_1,\ldots,2d_n$. Each of these subpaces will be called a ``shell'', and the set $\{V_1, \ldots, V_n\}$ a ``shell partition''. In quantum chemistry, the $V_i$'s can be the vector spaces spanned by sets of spin-orbital pairs (i.e. the spin-orbitals of opposite spin corresponding to the same atomic orbital). In such a case,  the  shells will be termed tentatively ``primitive shells'', as all the $d_i$'s are equal to $1$. Even with this restriction, there will be infinitely many possible shell partitions, as soon as $n>1$. A natural shell partition with larger values for some $d_i$'s occurs when the system has degenerate orbitals. If the sets of degenerate orbitals in increasing energy order are $d_1-,\ldots,d_n-$fold degenerate, then, the shells $V_i$'s can be defined as the $2d_i$-dimensional vector spaces spanned by the associated  degenerate pairs of spin-orbitals.\\

We denote by $u_i$ the single determinantal function built from a set of $2d_i$ normalized spin-orbitals,  
$\left( \chi_{i,1},...,\chi_{i,2d_i}\right)_{i=1,\ldots,n}$ spanning the shell $V_i$:
\begin{equation}
 \forall i\in\{1,\ldots ,n\},\ u_i= \chi_{i,1}\wedge\cdots\wedge\chi_{i,2d_i}\quad,
\end{equation}
where $\wedge$ is the Grassmann (or exterior) product (which is intrinsically antisymmetrical) \cite{Cassam94-jmc,Cassam03-jmp}. Note that choosing a different set of linearly independent spin-orbitals would only change $u_i$ by a constant factor.

\textbf{Remark:} In quantum chemistry, as mentionned above, the even dimension $2d_i$ of the $V_i$'s would arise from the fact that there are as many basis spin-orbitals of spin $+\frac{1}{2}$,  as there are of opposite spin. However, in the following, to alleviate notation, we will not distinguish the spin of the spin-orbitals. In other words, the spin-orbitals will be labelled by indices running from $1$ to $2d_i$, irrespective of their spin.\\

The symbol $(\chi)_1$ will designate the concatenated bases of the $n$ shells,
\begin{equation}
(\chi)_1:= (\chi_{1,1},\ldots,\chi_{1,2d_1},\chi_{2,1},\ldots,\chi_{2,2d_2},\ldots,\chi_{n,1},\ldots,\chi_{n,2d_n})\quad,
\end{equation}
which is a basis of the one-particle Hilbert space $V$. We will further denote by $(\chi)_N$ the $N$-particle basis set of Slater determinants induced by $(\chi)_1$
\begin{equation}
(\chi)_N:= (\chi_{i_1,j_1}\wedge\ldots\wedge\chi_{i_N,j_N})_{(i_1,j_1)<\cdots<(i_N,j_N)}\quad,
\end{equation}
where the order on the ordered pairs is the lexicographic order: $(i,j)<(k,l)$ if $i<k$ or if $i=k$ and $j<l$.
The union of all these basis sets, $(\chi):=\bigcup\limits_i (\chi)_i$,  including $(\chi)_0:=(1)$, is a basis of the first quantization equivalent of the Fock space.\\

In second quantization, the $\left( \chi_{i,1},\ldots,\chi_{i,2d_i}\right)$'s are created by the operators  $a^\dagger_{i,1},\ldots,a^\dagger_{i,2d_i}$, respectively, acting on the empty state $|0\rangle$:
\begin{equation}
 a^\dagger_{i,j}|0\rangle = |\chi_{i,j}\rangle\quad,
 \label{creation}
\end{equation}
so that, 
\begin{equation}
a^\dagger_{i,1}\cdots a^\dagger_{i,2d_i}|0\rangle = |\chi_{i,1}\wedge\cdots\wedge\chi_{i,2d_i}\rangle = |u_i\rangle \quad.
\end{equation}

Since the basis $(\chi)$ is not necessarily orthogonal, the corresponding annihilation operators $a_{i,j}$'s, defined by conjugation from Eq~\eqref{creation}:
$\langle0|a_{i,j} = \langle\chi_{i,j}|$ are not very convenient, because $ \langle0|a_{i,j}a^\dagger_{k,l}|0\rangle=\langle\chi_{i,j}|\chi_{k,l} \rangle \neq \delta_{(i,j),(k,l)}$.
In consequence, we introduce the dual basis $(\widetilde{\chi})$, that is the unique basis verifying the following property:
\begin{equation}
 \forall {i,j,k,l}, \ \langle\widetilde{\chi}_{i,j}|\chi_{k,l} \rangle = \delta_{(i,j),(k,l)} \quad,
\end{equation}
where $\delta_{(i,j),(k,l)}$ is the Kr\"onecker symbol for the ordered pair indices $(i,j)$ and $(k,l)$. 
The corresponding annihilation operators, denoted by a tilde, that is to say:  $\langle\widetilde{\chi}_{i,j}| = \langle0|\widetilde{a}_{i,j}$, satisfy the desired relationship:
\begin{equation}
\langle0|\widetilde{a}_{i,j}a^\dagger_{k,l}|0\rangle= \delta_{(i,j),(k,l)}\quad .
\end{equation}

It is also convenient to extend the notion of creation and annihilation operators to arbitrary quantum states. So, we define the creation operator, $a^\dagger(f)$, of a general state, $f=\sum\limits_{(i_1,j_1),\ldots,(i_k,j_k)} c_{(i_1,j_1),\ldots,(i_k,j_k)} \, \chi_{i_1,j_1}\wedge\cdots\wedge\chi_{i_k,j_k}$, $c_{(i_1,j_1),\ldots,(i_k,j_k)}\in \mathbb{C}$, as follows:
\begin{equation}
a^\dagger(f)|0\rangle=|f\rangle
=\sum\limits_{(i_1,j_1),\ldots,(i_k,j_k)} c_{(i_1,j_1),\ldots,(i_k,j_k)}\, a^\dagger_{i_1,j_1}\cdots a^\dagger_{i_k,j_k}|0\rangle\quad.
\end{equation}

For example, $a^\dagger(\chi_{i,j})=a^\dagger_{i,j}$ and $a^\dagger(u_i)=a^\dagger_{i,1}\cdots a^\dagger_{i,2d_i}$.\\

We define the ``dual'' annihilation operator of a product state, $\widetilde{a}(u_i)$, as the product of the dual annihilation operators, $\widetilde{a}(\chi_{i,j})=\widetilde{a}_{i,j}$, in reverse order: $\widetilde{a}(u_i)=\widetilde{a}_{i,2d_i}\cdots \widetilde{a}_{i,1}$, and more generally, by anti-linearity,
the ``dual'' annihilation operator of $a^\dagger(f)$ as
\begin{equation}
\widetilde{a}(f)=\sum\limits_{(i_1,j_1),\ldots,(i_k,j_k)} \bar{c}_{(i_1,j_1),\ldots,(i_k,j_k)} \, \widetilde{a}_{i_k,j_k}\cdots \widetilde{a}_{i_1,j_1}\quad,
\end{equation}
where the bar $\bar{c}$ denotes complex conjugation.

\textbf{Definition:} 
We say that a $(2d_i-k)$-particle Slater determinant \textbf{\emph{$x$ is included in $u_i$}} if there exists a set $\{h_1,\ldots,h_k\}$ such that $a^\dagger(u_i)=a^\dagger_{i,h_1}\cdots a^\dagger_{i,h_k} a^\dagger(x)$.

So, for every Slater determinant $m\in (\chi)_N$ of the $N$-particle induced basis set, 
we can write:
\begin{equation}
  a^\dagger(m) = a^\dagger(u_{i_1})\cdots a^\dagger(u_{i_\omega}) a^\dagger(x_{j_1})\cdots a^\dagger(x_{j_\Omega})\quad,
  \label{qr-factorization}
\end{equation}
where the $x_{j_k}$'s are strictly included in some $u_{j_k}$'s which are distinct from one another and from $u_{i_1},\ldots,u_{i_\omega}$.

\textbf{Definition:}
We call $\omega$ the \textbf{\emph{generalized ionicity}} of $m$ in the $u_i$'s. It represents the number of fully occupied shells. 
Note that it is called the \emph{degree of $m$ in the $u_i$'s} in mathematics \cite{Vivier56}. When the shells are chosen to be a set of primitive shells, $\omega$ is the ionicity number of the Slater determinant $m$, as defined in \cite{Simonetta1968} in the context of VB wave functions. 
 
\textbf{Definition:}
The integer $\Omega$ is called the \textbf{\emph{generalized seniority number}} of $m$ relative to the $u_i$'s. It represents the number of non-empty, non-fully occupied shells.
Note that, when the shells are chosen to be a set of primitive shells, $\Omega$ is nothing but the seniority number of the Slater determinant $m$. 
 
\textbf{Remark:}
The integer $p=2\omega+\Omega$ is called the \textbf{\emph{reduced degree}} of $m$. It coincides with the number of particles of the Slater determinant in the primitive shell case i.e. when $d_1=\ldots=d_n=1$ (since, in this case, the $u_i$'s are $2$-particle states and the $x_j$'s are necessarily $1$-particle states).

The vector space spanned by all the Slater determinants, $m$, of the same GSN $\Omega$ is noted $M(\Omega)$.
It only depends upon the shell partition and not upon the choice of the shell basis sets. 
By extension, all wave functions in subspace $M(\Omega)$ will be said of GSN $\Omega$.
The subset of $M(\Omega)$ containing the wave functions spanned by Slater determinants $m$ of the same GI $\omega$ in the $u_i$'s is a subvector space of $M(\Omega)$, noted $M(\omega,\Omega)$ with $\omega\in\{0, \ldots , n-\Omega\}$. For a given $\Omega$,  $M(\Omega)$ is the direct sum of all the $M(\omega,\Omega)$'s. The $M(\omega,\Omega)$'s can be further decomposed into their projections onto the $N$-particle Hilbert spaces, noted $M(N,\omega,\Omega)$. In the next section, we will introduce a GSN operator, which acts diagonally on the $M(\Omega)$'s and whose expectation value on a normalized element of each $M(\Omega)$ is its GSN.\\

\section{Hermitian operators related to the GSN and GI concepts}

\textbf{Definition:}
For  $i\in\{1,\ldots,n\}$ and any quantum state $F$,  we consider the decomposition: 
\begin{equation}
 a^\dagger(F)=\hat{Q}_i(F)+\hat{R}_i(F)\quad,
 \label{QR-decomp}
\end{equation}
where $\hat{Q}_i(F)$ represents the  part of the $a^\dagger(F)$'s expansion in the $(\chi)$-basis containing at least one $a^\dagger_{i,j}$, and $\hat{R}_i(F)$ the part of $a^\dagger(F)$ which does not contain any creation operator of a spin-orbital appearing in $u_i$. We call it the \textbf{\emph{residue}} or the \textbf{\emph{rest}} of $F$ relatively to $u_i$ in the basis $(\chi)$.\\
$\hat{R}_i(F)$ can be expressed as
\begin{equation}
\hat{R}_i(F)=\widetilde{a}(u_i)a^\dagger(u_i)a^\dagger(F)\quad,
 \label{R-def}
\end{equation}
 and $\hat{Q}_i(F)$ can be further decomposed as
\begin{equation}
\hat{Q}_i(F)=\mathring{Q}_i(F)+a^\dagger(u_i)\widetilde{a}(u_i)a^\dagger(F)\quad,
 \label{Q-decomp}
\end{equation}
where $\mathring{Q}_i(F)$ represents the part of the $a^\dagger(F)$'s expansion containing at least one $a^\dagger_{i,j}$ but not $a^\dagger(u_i)$ entirely.\\
By combining Eqs. \eqref{QR-decomp}, \eqref{R-def} and \eqref{Q-decomp}, we obtain,
\begin{equation}
\mathring{Q}_i(F)=\Big(1-\widetilde{a}(u_i)a^\dagger(u_i)-a^\dagger(u_i)\widetilde{a}(u_i)\Big)a^\dagger(F).
\label{Q-ring}
\end{equation}

\textbf{Remark:}
More generally, we can define $\mathring{Q}_{i_1,i_2,\ldots,i_k}(F)=\mathring{Q}_{i_1}\mathring{Q}_{i_2}\cdots\mathring{Q}_{i_k}(F)$, (where the order of the $i_j$'s is indifferent since the $\mathring{Q}_{i_j}$'s commute), which extracts the  part of the $a^\dagger(F)$'s expansion containing at least one $a^\dagger_{i_1,j_1}$, one $a^\dagger_{i_2,j_2}$, ... and one $a^\dagger_{i_k,j_k}$, without containing entirely $a^\dagger(u_{i_1})$ nor $a^\dagger(u_{i_2})$ nor ... nor $a^\dagger(u_{i_k})$.\\

\textbf{Definition:}
The linear operator $\hat{\Omega} : \ G \longmapsto \hat{\Omega}(G):=\sum\limits_{i=1}^n \mathring{Q}_i(G)$ is called the \textbf{\emph{generalized seniority number operator}}. It acts diagonally on any element of $M(\Omega)$:
\begin{equation}
  \forall G\in M(\Omega),  \quad\hat{\Omega}(G) = \ \Omega\,a^\dagger(G)  \quad.
  \label{Omega-id}
\end{equation}
To prove the latter identity, let $G\in M(\Omega)$. The creation operator
$a^\dagger(G)$ can be regarded as a linear combination of $a^\dagger(m)$'s, with $m\in(\chi)$ the induced basis of Slater determinants. For all $m$, we can write $a^\dagger(m)=a^\dagger(u_{i_1})\cdots a^\dagger(u_{i_\omega}) a^\dagger(x_{j_1})\cdots a^\dagger(x_{j_\Omega})$ (for some $\omega$-value) with $a^\dagger(x_{j_1})\cdots a^\dagger(x_{j_\Omega})|0\rangle\in M(0,\Omega)$. Applying $\hat{\Omega}$ to 
$m$ and using Eq. \eqref{Q-ring}, the only non zero contributions come from $\mathring{Q}_{j_1}(m)=a^\dagger(m),\ldots,\mathring{Q}_{j_\Omega}(m)=a^\dagger(m)$, respectively. So, we find  exactly $\Omega$ times $a^\dagger(m)$ in $\hat{\Omega}(m)$.
This being true for all the $a^\dagger(m)$'s appearing in the expression of $a^\dagger(G)$, by linearity of $\hat{\Omega}$, we obtain the identity, Eq. \eqref{Omega-id}.\\

Similarly, a generalized ionicity operator can be defined as follows:\\
\textbf{Definition:}
The linear operator 
  $\hat{\omega} : \ G \longmapsto \hat{\omega}(G) = \sum\limits_{i=1}^n a^\dagger(u_i)\widetilde{a}(u_i)a^\dagger(G)$
 is called the \textbf{\emph{generalized ionicity operator}} for the shell partition $\{V_1, \ldots, V_n\}$.
It does not depend upon a change of basis of $V_i$, for any $i$. It acts diagonally on any element of the $M(\omega,\Omega)$'s:
\begin{equation}
  \forall G\in M(\omega,\Omega),  \quad \hat{\omega}(G) = \ \omega\,a^\dagger(G)  \quad.
  \label{omega-id}
\end{equation}

To prove the latter identity, let $G\in M(\omega,\Omega)$. Applying $\hat{\omega}$ to an $m\in(\chi)$  in the expansion of $G$, whose creation operator can necessarily be cast in the form given in Eq. \eqref{qr-factorization},
 the only contributing terms are \mbox{$a^\dagger(u_{i_1})\widetilde{a}(u_{i_1})a^\dagger(m)=a^\dagger(m),\ldots,a^\dagger(u_{i_\omega})\widetilde{a}(u_{i_\omega})a^\dagger(m)=a^\dagger(m)$}, as  $a^\dagger(u_i)\widetilde{a}(u_i)a^\dagger(m)=0$, for all $i\notin\{i_1,...,i_\omega\}$. So, $a^\dagger(m)$ appears exactly $\omega$ times in  $\hat{\omega}(m)$.
This being true for all the $a^\dagger(m)$'s appearing in the expression of $a^\dagger(G)$, by linearity of $\hat{\omega}$, we obtain the identity, Eq. \eqref{omega-id}.\\

\textbf{Remark:}
A third identity follows from the previous two, that is to say, from Eqs. \eqref{Omega-id} and \eqref{omega-id}:
\begin{equation}
  \forall G\in M(\omega,\Omega), \quad (n-\omega-\Omega)\,a^\dagger(G) = \sum_{i=1}^n \hat{R}_i(G) \quad,
  \label{id3}
\end{equation}
where the integer $(n-\omega-\Omega)$ is the number of empty shells in $G$.\\
Indeed, let $G\in M(\omega,\Omega)$. By using Eqs. \eqref{QR-decomp} and \eqref{Q-decomp}, we can decompose $a^\dagger(G)$ in $n$ different manners as follows:
\begin{equation}
  \forall i\in\{1,...,n\}, \ a^\dagger(G) = \mathring{Q}_i(G) + a^\dagger(u_i)\widetilde{a}(u_i)a^\dagger(G) + \hat{R}_i(G)\quad.
\end{equation}
By summing these $n$ equalities and using Eqs. \eqref{Omega-id} and \eqref{omega-id},  we obtain:
\begin{equation}
 n\,a^\dagger(G) = \Omega\,a^\dagger(G) + \omega\,a^\dagger(G) + \sum_{i=1}^n \hat{R}_i(G)\quad,
\end{equation}
hence the result, Eq. \eqref{id3}.\\

The operator $\hat{\Omega}$  can be used to decompose  the creation operator of an arbitrary quantum state $F$ onto the vector spaces of definite GSN, that is the $M(\Omega)$'s. This can be achieved by using L\"owdin projectors \cite{lowdin66}, for example. Let
$a^\dagger(F)=\sum\limits_{\Omega=0}^n a^\dagger(G_\Omega)$, where $G_\Omega\in M(\Omega)$. From Eq. \eqref{Omega-id}, we deduce,
\begin{equation}
\hat{\Omega}(F) = \sum\limits_{\Omega=0}^n \Omega\,a^\dagger(G_\Omega) \quad. 
  \label{omega-F}
\end{equation}
For all $\Omega\neq0$, 
we can extract the $\Omega\, a^\dagger(G_\Omega)$ component of this decomposition by projection,
\begin{equation}
 \Omega\, a^\dagger(G_\Omega) = \prod_{\substack{0\leq j \leq n \\ j\neq \Omega}} \frac{\hat{\Omega}(F)-j\, a^\dagger(F)}{\Omega-j} \quad.
\end{equation}
Then, the generalized seniority-zero part of $a^\dagger(F)$ can be obtained by difference,
\begin{equation}
 a^\dagger(G_0) = a^\dagger(F)-\sum\limits_{\Omega\neq 0} a^\dagger(G_\Omega)= a^\dagger(F)-\sum\limits_{\Omega\neq 0}\frac{1}{\Omega} \prod_{\substack{0\leq j \leq n \\ j\neq \Omega}} \frac{\hat{\Omega}(F)-j\, a^\dagger(F)}{\Omega-j} \quad.
\end{equation}\\

\section{Example: GSN for the C$_{60}$ ``$\pi$-electron'' system}

The primitive shell partition used to define seniority numbers in quantum chemistry stems from the fact that spin-orbitals
of the same spin are degenerate with respect to spin symmetry for the spin-free Hamiltonian usually considered. It is therefore a natural idea to 
take also into account spatial symmetry, that is to say, to partition the one-particle Hilbert space into subspaces closed with respect to both spin and spatial symmetry operation. The highest, finite group,  spatial symmetry  known in molecular system is the icosahedral symmetry. So, we will illustrate our generalization of the seniority number concept on the buckminsterfullerene C$_{60}$ molecule.

\begin{figure}[ht]

\begin{center}
\includegraphics*[scale=0.7]{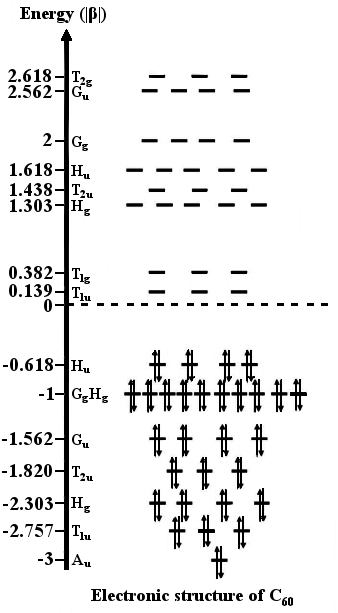}\\
\end{center}

\caption{Energy diagram of C$_{60}$ H\"uckel molecular orbitals with electron occupation in the ground state reference configuration. }

\label{C60-MOs}

\end{figure}

Although this molecule is not planar, we will consider that each carbon contributes one electron to a  ``$\pi$-electron'' system. Then,  at the Hückel level of theory, the one-particle Hilbert space is spanned by $60$ orbitals, so that $\dim V=120$.
If $V$ is partitionned into the corresponding $60$ primitive shells, we obtain the usual seniority numbers. However, even if one limits the CI space to seniority-zero Slater determinants, the latter will be of dimension ${60 \choose 30}\approx1.18\times10^{17}$, which is clearly untractable. So, to further decompose the seniority-zero space, we are going to use GSNs associated to the shells corresponding to the degenerate orbitals displayed in Fig. \ref{C60-MOs}.

More precisely, the shell partition consists of $15$ shells, $V_1, V_2, V_3, \cdots, V_{14}, V_{15}$ of dimensions $2, 6, 8, \cdots, 8 , 6$, respectively. The largest shell $V_6$ is of dimension $18$ due to an accidental degeneracy at the Hückel level. For any fixed GSN $\Omega$, the size of the $(\Omega)$-subspace of the seniority-zero space can be calculated by using basic combinatorics. The dimension of the $(\Omega=0)$-subspace is found to be $1464$, which corresponds to a small number of Slater determinants by modern standard. For  $\Omega\leq 1$, there are $601594$ additional Slater determinants to include, and for  $\Omega\leq 2$, an extra set of $53141130$ Slater determinants needs again to be added. All of these restricted CI subspaces should be amenable to quantum chemistry computations in contrast with the full seniority-zero space. However, GSN hierarchies should not be seen as an alternative but rather as a complement to hierarchies based on excitation level. In fact, these $(\Omega)$-subspace should be further refined into subspaces of fixed excitation numbers, so as to reduce further .\\

\section{Conclusion}

The concepts of GSN and GI, generalizing those of seniority number and ionicity, have been introduced with their associated operators.
The generalization is based on the partitionning of the one-particle Hilbert space into shells. From the mathematical point of view, the choice of the partition can be arbitrary. However, in practice, the partition should be chosen on physical ground. The GSN counts the number of partially occupied shells, whereas the GI counts the fully occupied shells.\\

In this paper, a partition of the spin-orbital basis functions according to their spatial and spin degeneracy has been illustrated on C$_{60}$.
A  hierarchy of CI-spaces based on the corresponding GSN  affords to split the seniority-zero
space of C$_{60}$ for a basis set of $60$ Hückel molecular orbitals into CI-subspaces of reduced dimensions,
lending themselves to numerical computations for low values of the GSN.\\

We suggest that generalized seniority numbers based on spatial symmetry can be relevant parameters to limit CI expansions, as already observed for seniority number.  This hypothesis relies on known phenomena, where a correlation has been established between the complete filling of a shell of a certain type and an unusual stability property. We have in mind the octet rule, the 18-electron rule or aromaticity, for example. Explorative ST0-3G/$\pi$-electron-FullCI calculations on bezene, $C_6H_6$, support this by the fact that the quadruple and hextuple excited configurations of highest weight in the CI-expansion, whose contributions are actually non-negligible, correspond to a GSN equal to zero. So,  in this case,  GSN would provide a mean to select the main hextuple excited configuration to take into account.\\

More generally, in a same way as seniority is useful for quantum systems exhibiting pairing phenomena, applications of GSN could be found in systems where a form of clustering occurs.\\ 

\section*{Acknowledgements}
This work was supported by the grant CARMA ANR-12-BS01-0017.\\

\newpage

\section*{Figures}

\end{document}